\documentstyle[ltwol,epsfig]{article}

\def\Journal#1#2#3#4{{#1} {\bf #2}, #3 (#4)}

\def\NPB{{\em Nucl. Phys.} B}
\def\PLB{{\em Phys. Lett.}  B}

\def\be{\begin{equation}}
\def\ee{\end{equation}}
\def\bea{\begin{eqnarray}}
\def\eea{\end{eqnarray}}

\newcommand{\gapprox}{\raisebox{-0.5ex}{$\
\stackrel{\textstyle>}{\textstyle\sim}\ $}}
\newcommand{\lsim}{\raisebox{-0.5ex}{$\
\stackrel{\textstyle<}{\textstyle\sim}\ $}}

\begin{document}
\title{THE FERMION MASS PROBLEM AND THE ANTI-GRAND
UNIFICATION MODEL}

\author{C. D. FROGGATT, M. GIBSON}

\address{Department of Physics and Astronomy, Glasgow University,
Glasgow G12 8QQ, Scotland}

\author{H. B. NIELSEN}

\address{Niels Bohr Institute, Blegdamsvej 17,
Copenhagen $\phi$, Denmark}

\author{D. J. SMITH}

\address{Institut f\"{u}r Physik, Humboldt Universit\"{a}t Berlin,
Invalidenstr. 110, 10115 Berlin, Germany}


\twocolumn[\maketitle\abstracts{ We describe the Anti-Grand
Unification Model (AGUT) and the Multiple Point Principle (MPP)
used to calculate the values of the Standard Model gauge coupling
constants in the theory, from the requirement of the existence
of degenerate vacua. The application of the MPP to the pure
Standard Model predicts the existence of a second minimum of the
Higgs potential close to the cut-off, which we take to be the Planck
scale, giving our Standard Model predictions for the top quark and
Higgs masses: $M_t = 173 \pm 5$ GeV and $M_H = 135 \pm 9$ GeV. We
also discuss mass protection by chiral charges and present a fit
to the charged fermion mass spectrum using the
chiral quantum numbers of the maximal AGUT gauge group
$SMG^3 \times U(1)_f$, where
$SMG \equiv SU(3) \times SU(2) \times U(1)$. The neutrino mass and
mixing problem is then briefly discussed for models with chiral
flavour charges responsible for the charged
fermion mass hierarchy.}]

\section{Introduction}
One of the outstanding problems in particle physics is to
explain the observed pattern of quark-lepton masses and
of flavour mixing. This is the problem of the hierarchy of
Yukawa coupling constants in the Standard Model (SM),
which range in value from of order 1 for the top quark
to of order $10^{-5}$ for the electron. However
there is no reason in the SM for the Higgs field to
prefer to couple to one fermion rather than
another; in fact one would expect them all to be
of order unity. We suggest \cite{fn} that the natural
resolution to this problem is the existence of some
approximately conserved chiral charges beyond the SM.
These charges, which we assume to be the gauge quantum numbers
in the fundamental theory beyond the SM, provide selection
rules forbidding the transitions
between the various left-handed and right-handed quark-lepton
states, except for the top quark. In order to generate mass
terms for the other fermion states,
we have to introduce new Higgs fields, which break the
fundamental gauge symmetry group $G$ down to the SM group.
We also need suitable intermediate fermion states to
mediate the forbidden transitions, which we take to be
vector-like Dirac fermions with a mass of order the
fundamental scale $M_F$ of the theory. In this way
effective SM Yukawa coupling constants are generated, which
are suppressed by the appropriate product of Higgs field
vacuum expectation values measured in units of $M_F$.

If we want to explain the observed spectrum of quarks and leptons,
it is clear that we need charges which---possibly in a
complicated way---separate the generations and, at least
for $t-b$ and $c-s$, also quarks in the same generation.
Just using the usual simple $SU(5)$ GUT charges does not
help, because both ($\mu_R$ and $e_R$) and
($\mu_L$ and $e_L$) have the same $SU(5)$ quantum numbers.
So we prefer to keep each SM irreducible representation
in a separate irreducible representation of $G$ and
introduce extra gauge quantum numbers distinguishing
the generations, by adding extra cross-product factors to
the SM gauge group. In this talk we consider the maximal
anomaly free gauge group of this type---the anti-grand
unification (AGUT) group $SMG^3\times U(1)_f$. In section 2
we discuss the structure of the AGUT model and the prediction
of the values of the SM gauge coupling constants, using the
so-called Multiple Point Principle (MPP). We apply this principle
to the pure SM in section 3, assuming a desert up to the
Planck scale, and obtain predictions for the top quark and SM
Higgs particle masses. In section 4 we consider the Higgs fields
responsible for breaking the AGUT gauge group and the structure
of the resulting quark-lepton mass matrices, together with details
of a fit to the observed spectrum. The problem of neutrino mass
and mixing in models with approximately conserved chiral flavour
charges are discussed in section 5. Finally we present our
conclusions in section 6.

\section{Anti-Grand Unification}

In the AGUT model the SM gauge group is extended in much the
same way as Grand Unified $SU(5)$ is often assumed; it is just
that we assume another non-simple gauge group
$G = SMG^3 \times U(1)_f$, where
$SMG \equiv SU(3) \times SU(2) \times U(1)$, becomes active
near the Planck scale $M_{Planck} \simeq 10^{19}$ GeV. So we
have a pure SM desert, without any supersymmetry,
up to an order of magnitude or so below $M_{Planck}$.
The existence of the $SMG^3 \times U(1)_f$ group means
that, near the Planck scale, each of the three quark-lepton
generations has got its own gauge group and associated
gauge particles with the same structure as the SM gauge group.
There is also an extra abelian $U(1)_f$ gauge boson, giving
altogether $3 \times 8 = 24$ gluons, $3 \times 3 = 9$ $W$'s and
$3 \times 1 + 1 =4$ abelian gauge bosons.

At first sight, this $SMG^3 \times U(1)_f$ group with
its 37 generators seems to be just one among many
possible SM gauge group extensions. However, it is
actually not such an arbitrary choice, as it
can be uniquely specified by postulating 4 reasonable
requirements on the gauge group $G \supseteq SMG$:

\begin{enumerate}
\item $G$ should transform the presently known (left-handed,
say) Weyl particles into each other, so that
$G \subseteq U(45)$. Here $U(45)$ is the group of all
unitary transformations of the 45 species of Weyl fields (3
generations with 15 in each) in the SM.
\item No anomalies, neither gauge nor mixed.
We assume that only straightforward anomaly
cancellation takes place and, as in the SM itself,
do not allow for a Green-Schwarz type anomaly
cancellation \cite{green-schwarz}.
\item The various irreducible representations of Weyl fields
for the SM group remain irreducible under $G$.
\item $G$ is the maximal group satisfying the other 3
postulates.
\end{enumerate}

With these four postulates a somewhat complicated
calculation shows that,
modulo permutations of the various irreducible representations
in the Standard Model
fermion system, we are led to our gauge group
$SMG^3\times U(1)_f$.
Furthermore it shows that the SM group is embedded
as the diagonal subgroup of $SMG^3$, as required
in our AGUT model. The AGUT group breaks
down an order of magnitude or so below the Planck
scale to the SM group. The anomaly cancellation constraints
are so tight that, apart from various permutations of the
particle names, the $U(1)_f$ charge assignments are
uniquely determined up to an overall normalisation and
sign convention. In fact the $U(1)_f$ group does not couple to
the left-handed particles or any first generation particles,
and the $U(1)_f$ quantum numbers can be chosen as follows:
\begin{equation}
Q_f(\tau_R) = Q_f(b_R) = Q_f(c_R) = 1
\end{equation}
\begin{equation}
Q_f(\mu_R) = Q_f(s_R) = Q_f(t_R) = -1
\end{equation}

\begin{figure}
\leavevmode
\centerline{\epsfig{file=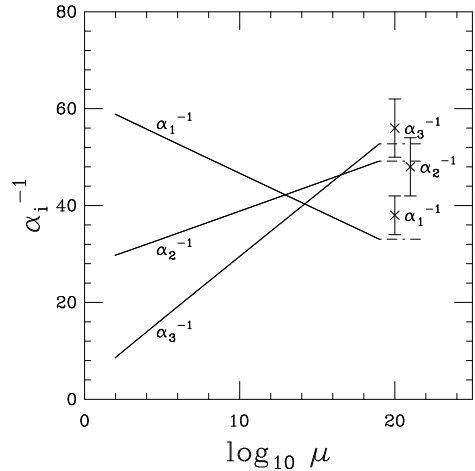,width=6.5cm}
}
\caption{Evolution of the Standard Model
fine structure constants $\alpha_i$ ($\alpha_1$
in the SU(5)
inspired normalisation) from the electroweak scale
to the Planck scale. The anti-GUT model
predictions for the values at the Planck scale,
$\alpha_i^{-1}(M_{Planck})$,
are shown with error bars.}
\label{fig:alphas}
\end{figure}

The SM gauge coupling constants do not,
of course, unify in the AGUT model, but their values have
been successfully calculated using the Multiple Point
Principle \cite{glasgowbrioni}. According to the MPP, the
coupling constants should be fixed such as to ensure the
existence of many vacuum states with the same energy density;
in the Euclideanised version of the theory, there is a
corresponding phase transition. So if several vacua
are degenerate, there is a multiple point. The couplings
at the multiple points have been calculated in
lattice gauge theory for the groups $SU(3)$,
$SU(2)$ and $U(1)$ separately. We imagine that the
lattice has a truly physical significance in providing
a cut-off for our model at the Planck scale. The SM
fine structure constants correspond to those of
the diagonal subgroup of the $SMG^3$ group and,
for the non-abelian groups, this gives:
\begin{equation}
\alpha_i(M_{Planck}) = \frac{\alpha_i^{Multiple \ Point}}{3}
\qquad i=2,\ 3
\end{equation}
The situation is more complicated for the abelian
groups, because it is possible to have gauge invariant
cross-terms between the different $U(1)$ groups in
the Lagrangian density such as:
\begin{equation}
\frac{1}{4g^2} F_{\mu\nu}^{gen \ 1}(x) F_{gen \ 2}^{\mu \nu}(x)
\end{equation}
So, in first approximation, for the SM $U(1)$ fine
structure constant we get:
\begin{equation}
\alpha_1(M_{Planck}) = \frac{\alpha_1^{Multiple \ Point}}{6}
\end{equation}
The agreement of these AGUT predictions with the data
is shown in figure 1.

\section{The MPP Prediction for the Top Quark and
Higgs masses in the Standard Model}

\begin{figure}
\leavevmode
\centerline{
\epsfig{file=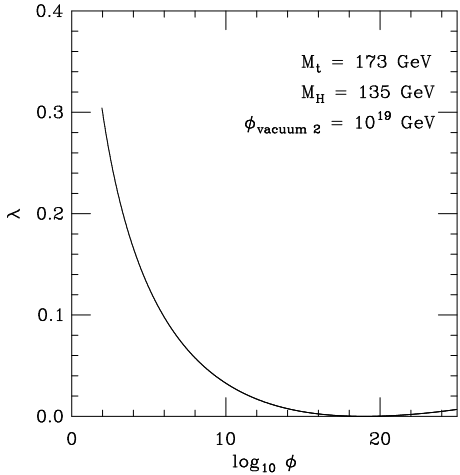,width=6.8cm}}
\caption{Plot of $\lambda$ as a function
of the scale of the
Higgs field $\phi$ for degenerate vacua with the second Higgs
VEV at the Planck scale $\phi_{vac\;2}=10^{19}$ GeV.}
\label{fig:lam19}
\end{figure}
The application of the MPP to the pure Standard Model (SM), with
a cut-off close to $M_{Planck}$, implies
that the SM parameters should be adjusted, such that there exists
another vacuum state degenerate in energy density with the
vacuum in which we live. This means that the effective SM
Higgs potential $V_{eff}(|\phi|)$
should, have a second minimum
degenerate with the well-known first
minimum at the electroweak scale
$\langle |\phi_{vac\; 1}| \rangle = 246$ GeV.
Thus we predict that our vacuum is barely stable and we
just lie on the vacuum stability curve in the top quark, Higgs
particle (pole) mass ($M_t$, $M_H$) plane.
Furthermore we expect the second minimum to be within an
order of magnitude or so of the fundamental scale,
i.e. $\langle |\phi_{vac\; 2}| \rangle \simeq M_{Planck}$.
In this way, we essentially select a particular point on
the SM vacuum stability curve and hence the MPP condition
predicts precise values for $M_t$ and $M_H$.

For the purposes of our discussion it is sufficient to consider
the renormalisation group improved tree level effective
potential $V_{eff}(\phi)$.
We are interested in values of the Higgs field
of the order $|\phi_{vac\; 2}| \simeq M_{Planck}$,
which is very large compared to the electroweak scale,
and for which the quartic term
strongly dominates the $\phi^2$ term;
so to a very good approximation
we have:
\begin{equation}
V_{eff}(\phi) \simeq
\frac{1}{8}\lambda (\mu = |\phi | ) |\phi |^4
\end{equation}
The running Higgs self-coupling constant $\lambda (\mu)$
and the top quark running Yukawa coupling constant $g_t(\mu)$
are readily computed by means of the
renormalisation group equations, which
are in practice solved numerically, using the second order
expressions for the beta functions.

\begin{figure}
\leavevmode
\centerline{
\epsfig{file=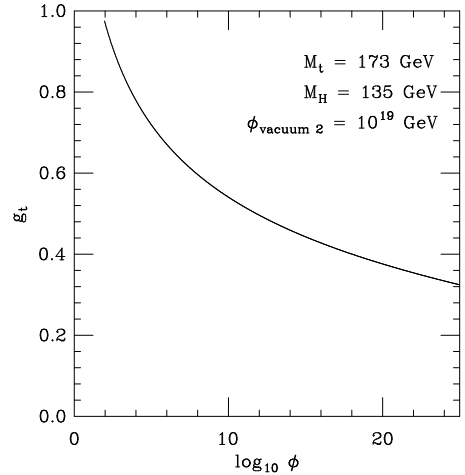,width=6.8cm}
}
\vspace{-0.6cm}
\caption{Plot of $g_t$ as a function
of the scale of the
Higgs field $\phi$ for degenerate vacua with the second Higgs
VEV at the Planck scale $\phi_{vac\;2}=10^{19}$ GeV.}
\label{fig:top19}
\end{figure}

The vacuum degeneracy condition is imposed by requiring:
\begin{equation}
V_{eff}(\phi_{vac\; 1}) = V_{eff}(\phi_{vac\; 2})
\label{eqdeg}
\end{equation}
Now the energy density in vacuum 1 is exceedingly small
compared to $\phi_{vac\; 2}^4 \simeq M_{Planck}^4$. So
we basically get the degeneracy condition, eq.~(\ref{eqdeg}),
to mean that the coefficient $\lambda(\phi_{vac\; 2})$
of $\phi_{vac\; 2}^4$ must be zero with high accuracy.
At the same $\phi$-value the derivative of the effective
potential $V_{eff}(\phi)$ should be zero, because it has
a minimum there.
Thus at the second minimum of the effective potential
the beta function $\beta_{\lambda}$ also vanishes:
\begin{equation}
\beta_{\lambda}(\mu = \phi_{vac\; 2}) =
\lambda(\phi_{vac\; 2}) = 0
\end{equation}
which gives to leading order the relationship:
\begin{equation}
\frac{9}{4}g_2^4 + \frac{3}{2}g_2^2g_1^2 +
\frac{3}{4}g_1^4 - 12g_t^4 = 0
\end{equation}
between the top quark Yukawa coupling and the electroweak
gauge coupling constants $g_1(\mu)$ and $g_2(\mu)$
at the scale $\mu = \phi_{vac\; 2} \simeq M_{Planck}$.
We use the renormalisation group equations to relate the couplings
at the Planck scale to their values
at the electroweak scale.
Figures \ref{fig:lam19} and \ref{fig:top19} show
the running coupling constants $\lambda(\phi)$ and $g_t(\phi)$
as functions of $\log(\phi)$. Their values at the
electroweak scale give our predicted combination of pole
masses \cite{fn2}:
\begin{equation}
M_{t} = 173 \pm 5\ \mbox{GeV} \quad M_{H} = 135 \pm 9\ \mbox{GeV}
\end{equation}

\section{Fermion Mass Hierarchy in AGUT}

The $SMG^3 \times U(1)_f$ gauge group is broken by a set of
Higgs fields $S$, $W$, $T$ and $\xi$ down to the SM gauge
group. Together with the Weinberg Salam Higgs field,
$\phi_{WS}$, they are responsible for breaking the quark-lepton
mass protection by the chiral AGUT quantum numbers. We have
the freedom of choosing the abelian quantum numbers of the Higgs
fields, which we can express as charge vectors of the form:
\begin{equation}
\vec{Q} \equiv \left ( \frac{y_1}{2},\frac{y_2}{2},
\frac{y_3}{2},Q_f \right ),
\end{equation}
where $y_i/2$ $(i= 1, 2, 3)$ are the $U(1)_i$ weak hypercharges.
However we fix their non-abelian representations by imposing
a natural generalisation of the SM charge quantisation rule
\begin{equation}
y_i/2 + d_i /2 + t_i/3 = 0 \quad ( \mbox{mod} \quad 1)
\end{equation}
and requiring that they be singlet or fundamental representations.
The duality, $d_i$, and triality, $t_i$, here are given by
$d_i = +1, 0$ for the doublet and singlet representations respectively
of $SU(2)_i$,
and $t_i = +1, -1, 0$ for the $\boldmath{3, \overline{3}, 1}$
representations of $SU(3)_i$.

By requiring a realistic charged fermion spectrum
(with $\phi_{WS}$ giving an unsuppressed top quark mass),
we are led to the following choice:
\begin{eqnarray}
\vec{Q}_{\phi_{WS}}  =  (0, \frac{2}{3}, -\frac{1}{6}, 1)&&,\quad
\vec{Q}_W  =  (0, -\frac{1}{2}, \frac{1}{2}, -\frac{4}{3}), \nonumber\\
\vec{Q}_T   =   (0, -\frac{1}{6}, \frac{1}{6}, -\frac{2}{3})&&,\quad
\vec{Q}_{\xi}   =   (\frac{1}{6}, -\frac{1}{6}, 0, 0), \nonumber\\
\vec{Q}_S   =   (\frac{1}{6}, -\frac{1}{6}, 0 , -1)&&
\end{eqnarray}
The orders of magnitude
for the effective SM Yukawa coupling matrix
elements are then given by:
\begin{eqnarray}
Y_U  & \simeq &  \left ( \begin{array}{ccc}
	S^{\dagger}W^{\dagger}T^2(\xi^{\dagger})^2 & W^{\dagger}T^2\xi &
		(W^{\dagger})^2T\xi \\
	S^{\dagger}W^{\dagger}T^2(\xi^{\dagger})^3 & W^{\dagger}T^2 &
		(W^{\dagger})^2T \\
	S^{\dagger}(\xi^{\dagger})^3 & 1 & W^{\dagger}T^{\dagger}
			\end{array} \right ),\\
Y_D  & \simeq & \left ( \begin{array}{ccc}
	SW(T^{\dagger})^2\xi^2 & W(T^{\dagger})^2\xi & T^3\xi \\
	SW(T^{\dagger})^2\xi & W(T^{\dagger})^2 & T^3 \\
	SW^2(T^{\dagger})^4\xi & W^2(T^{\dagger})^4 & WT
			\end{array} \right ),\\
Y_E  & \simeq &  \left (\hspace{-0.1cm}\begin{array}{ccc}
	SW(T^{\dagger})^2\xi^2 & W(T^{\dagger})^2(\xi^{\dagger})^3 &
		(S^{\dagger})^2WT^4\xi^{\dagger} \\
	SW(T^{\dagger})^2\xi^5 & W(T^{\dagger})^2 &
	(S^{\dagger})^2WT^4\xi^2 \\
	S^3W(T^{\dagger})^5\xi^3 & (W^{\dagger})^2T^4 & WT
			\end{array}\hspace{-0.1cm}\right)
\end{eqnarray}
Here $W, T, \xi, S$ should be interpreted as
the vacuum expectation values (VEVs) of
the Higgs fields in units of $M_{Planck}$
We have used the Higgs fields $W, T, \xi, S$
and the fields $W^{\dagger}, T^{\dagger}, \xi^{\dagger}, S^{\dagger}$
(with opposite charges) equivalently here, which we
can do in non-supersymmetric models.
In our fit below we do not need any suppression from the Higgs
field $S$ and so we set its VEV to be $S = 1$. This means that
the quantum numbers of the other Higgs fields $\vec{Q}_{\phi_{WS}}$,
$\vec{Q}_{W}$, $\vec{Q}_T$ and $\vec{Q}_{\xi}$ are only
determined modulo $\vec{Q}_S$.

Since the diagonals of $Y_U$, $Y_D$ and $Y_E$
are equal we expect to have the approximate relations
\begin{equation}
m_b \approx m_{\tau}, \quad m_s \approx m_{\mu}
\end{equation}
at $M_{Planck}$, since these masses come from the diagonal elements.
There are no such relations involving the top or charm quark masses,
since they come from off-diagonal elements which dominate
$Y_U$. We also note that we expect
\begin{equation}
m_d \gapprox m_u \approx m_e
\end{equation}
at $M_{Planck}$, since there are two approximately equal contributions
to the down quark mass.

The VEVs of the three Higgs fields $W$, $T$ and $\xi$
are taken to be free parameters in a
fit \cite{fgns} to the 12 experimentally known charged
fermion masses and mixing angles. The results of the best
fit, which reproduces all the experimental data to
within a factor of 2, are given in table \ref{convbestfit}
and correspond to the parameters
\begin{table}
\caption{Best fit to experimental data.
All masses are running
masses at 1 GeV except the top quark mass which is the pole mass.}
\begin{displaymath}
\begin{array}{|ccc|}
\hline
 & {\rm Fitted} & {\rm Experimental} \\ \hline
m_u & 3.6 {\rm \; MeV} & 4 {\rm \; MeV} \\
m_d & 7.0 {\rm \; MeV} & 9 {\rm \; MeV} \\
m_e & 0.87 {\rm \; MeV} & 0.5 {\rm \; MeV} \\
m_c & 1.02 {\rm \; GeV} & 1.4 {\rm \; GeV} \\
m_s & 400 {\rm \; MeV} & 200 {\rm \; MeV} \\
m_{\mu} & 88 {\rm \; MeV} & 105 {\rm \; MeV} \\
M_t & 192 {\rm \; GeV} & 180 {\rm \; GeV} \\
m_b & 8.3 {\rm \; GeV} & 6.3 {\rm \; GeV} \\
m_{\tau} & 1.27 {\rm \; GeV} & 1.78 {\rm \; GeV} \\
V_{us} & 0.18 & 0.22 \\
V_{cb} & 0.018 & 0.041 \\
V_{ub} & 0.0039 & 0.0035 \\ \hline
\end{array}
\end{displaymath}
\label{convbestfit}
\end{table}
\begin{equation}
\langle W\rangle  =  0.179 \quad
\langle T\rangle  =  0.071  \quad
\langle \xi\rangle  =  0.099 \label{WTxivev},
\end{equation}
in Planck units.
This fit is as good as we can expect in a model making order
of magnitude predictions.

\section{Neutrino Mass and Mixing Problem}

There is now strong evidence that the neutrinos are not massless as
they would be in the SM. Physics beyond the SM
can generate an effective light neutrino mass term
\begin{equation}
{\cal L}_{\nu-mass} = \sum_{i, j} \psi_{i\alpha}
\psi_{j\beta} \epsilon^{\alpha \beta} (M_{\nu})_{ij}
\end{equation}
in the Lagrangian, where $\psi_{i, j}$ are the Weyl spinors
of flavour $i$ and $j$, and $\alpha, \beta = 1, 2$.
Fermi-Dirac statistics mean that the mass matrix $M_{\nu}$
must be symmetric.

In models with chiral flavour symmetry we typically expect the elements
of the mass matrices to have different orders of magnitude. The charged
lepton matrix is then expected to give only a small contribution
to the lepton mixing. As a result of the symmetry of the neutrino mass
matrix and the hierarchy of the mass matrix elements it is natural
to have an almost degenerate pair of neutrinos, with
nearly maximal mixing\cite{degneut}. This occurs when an off-diagonal
element dominates the mass matrix.

The recent Super-Kamiokande data on the atmospheric neutrino anomaly
strongly suggests large $\nu_{\mu}-\nu_{\tau}$ mixing
with $\Delta m^2_{\nu_{\mu} \nu_{\tau}} \sim 10^{-3}$ eV$^2$.
Large $\nu_{\mu} - \nu_{\tau}$ mixing is given by the mass matrix
\begin{equation}
M_{\nu} =
\left(
\begin{array}{ccc}\times & \times & \times \\
\times & \times & A \\
\times & A & \times \end{array}\right )
\label{Mnu1}
\end{equation}
and we have
\begin{eqnarray}
& &\Delta m^2_{23} \ll \Delta m^2_{12} \sim \Delta m^2_{13}\\
& &\sin^2 \theta_{23} \sim 1
\end{eqnarray}
However, this hierarchy in $\Delta m^2$'s is inconsistent with
the small angle (MSW) solution to the solar neutrino problem,
which requires $\Delta m^2_{12} \sim 10^{-5} \ \mbox{eV}^2$.

Hence we need extra structure for the mass matrix such as having
several elements of the same order of magnitude. {\em e.g.}
\begin{equation}
M_{\nu} =
\left(
\begin{array}{ccc}a & A & B\\
A & \times & \times \\
B & \times & \times \end{array}\right )
\label{Mnu2}
\end{equation}
with $A \sim B \gg a$.
This gives
\begin{equation}
\frac{\Delta m^2_{12}}{\Delta m^2_{23}} \sim \frac{a}{\sqrt{A^2 + B^2}}.
\end{equation}
The mixing is between all three flavours. and is given by
the mixing matrix
\begin{equation}
U_{\nu} \sim \left( \begin{array}{ccc}
\frac{1}{\sqrt{2}} & -\frac{1}{\sqrt{2}} & 0\\
\frac{1}{\sqrt{2}} \cos \theta & \frac{1}{\sqrt{2}} \cos \theta &
        -\sin \theta\\
\frac{1}{\sqrt{2}} \sin \theta & \frac{1}{\sqrt{2}} \sin \theta &
        \cos \theta\\
\end{array} \right)
\end{equation}
where $\theta = \tan^{-1} \frac{B}{A}$.
So we have large $\nu_{\mu} - \nu_{\tau}$ mixing with
$\Delta m^2 = \Delta m^2_{23}$, and nearly maximal electron neutrino
mixing with $\Delta m^2 = \Delta m^2_{12}$.
However the AGUT model naturally gives a structure like
eq. (\ref{Mnu1}) rather than eq. (\ref{Mnu2}).

There is also some difficulty in obtaining the required
mass scale for the neutrinos. In models such as the AGUT the neutrino
masses are generated via super-heavy intermediate fermions in
a see-saw type mechanism. This leads to too small neutrino masses:
\begin{equation} m_{\nu} \lsim \frac{\langle{\phi_{WS}} \rangle^2}{M_F}
\sim 10^{-5}\ \mbox{eV},
\end{equation}
for $M_F = M_{Planck}$ (in general $m_{\nu}$ is also supressed
by the chiral charges). So we need to introduce a new mass scale
into the theory.
Either some intermediate particles with
mass $M_F \lsim 10^{15}\ \mbox{GeV}$, or an $SU(2)$ triplet
Higgs field $\Delta$ with
$\langle \Delta^0 \rangle \sim 1 \mbox{eV}$ is required.
Without further motivation the introduction of such particles
is {\em ad hoc}.

\section{Conclusions}

We presented two applications of the Multiple Point
Principle, according to which nature should choose
coupling constants such that the vacuum can exist
in degenerate phases. Applied to the AGUT model, it
successfully predicts the values of the three
fine structure constants, as illustrated in figure 1.
In the case of the pure SM, it leads to our predictions
for the top quark and Higgs pole masses:
$M_t = 173 \pm 5$ GeV and $M_H = 135 \pm 9$ GeV.

The maximal AGUT group $SMG^3 \times U(1)_f$
assigns a unique set of anomaly free chiral gauge
charges to the quarks and leptons. With an appropriate
choice of Higgs field quantum numbers, the AGUT chiral
charges naturally give a realistic charged fermion
mass hierarchy. An order of magnitude fit
in terms of 3 Higgs VEVs is
given in table \ref{convbestfit}, which
reproduces all the masses and mixing angles within a factor
of two. On the other hand, the puzzle of the neutrino masses
and mixing angles presents a challenge to the model.


\section*{References}
\begin{thebibliography}{99}

\bibitem{fn}C.D. Froggatt and H.B. Nielsen,
\Journal{\NPB}{147}{277}{1979}.

\bibitem{green-schwarz} M.B. Green and J. Schwarz,
\Journal{\PLB}{149}{117}{1984}.

\bibitem{glasgowbrioni}D.L. Bennett, C.D. Froggatt and H.B. Nielsen in
{\em Proceedings of the 27th
International Conference on High Energy Physics}
(Glasgow, 1994), eds.~P. Bussey and I. Knowles,
(IOP Publishing Ltd, 1995) p.~557;
{\em Perspectives in Particle Physics '94},
eds. D. Klabu\u{c}ar, I. Picek and D. Tadi\'{c},
(World Scientific, 1995) p.~255, hep-ph/9504294.

\bibitem{fn2}C.D. Froggatt and H.B. Nielsen,
\Journal{\PLB}{368}{96}{1996}.

\bibitem{fgns}
C.D. Froggatt, M. Gibson, H.B. Nielsen and D.J. Smith,
{\em Int. J. Mod. Phys.} A (to be published),
hep-ph/9706212.

\bibitem{degneut}C.D. Froggatt and H.B. Nielsen
\Journal{\NPB}{164}{114}{1979}

\end{thebibliography}
\end{document}